\documentclass[fleqn,usenatbib]{mnras}

\usepackage{newtxtext,newtxmath}
\usepackage[T1]{fontenc}
\usepackage{ae,aecompl}
\usepackage{url}

\usepackage{graphicx}	% Including figure files
\usepackage{amsmath}	% Advanced maths commands
\usepackage{amssymb}	% Extra maths symbols
\usepackage{float}

\newcommand{\nup}{$\nu_{\rm peak}^S$}

\usepackage{lineno}
\usepackage{ulem}
\title[TXS\,0506+056 is not a BL
  Lac]{TXS\,0506+056, the first cosmic neutrino source, is not a BL Lac}

\author[P. Padovani et al.]{P. Padovani$^{1,2}$\thanks{E-mail:
ppadovan@eso.org}, F. Oikonomou$^{1}$, M. Petropoulou$^3$, 
P. Giommi$^{4,5,6}$, E. Resconi$^7$ \\
$^{1}$European Southern Observatory, Karl-Schwarzschild-Str. 
2, D-85748 Garching bei M\"unchen, Germany\\
$^{2}$Associated to INAF - Osservatorio Astronomico di Roma, via Frascati 33,
I-00040 Monteporzio Catone, Italy\\
$^3$Department of Astrophysical Sciences, Princeton University, New Jersey 08544, USA
\\
$^{4}$Agenzia Spaziale Italiana, ASI, via del Politecnico s.n.c., I-00133 Roma Italy \\
$^{5}$Institute for Advanced Studies, Technische 
Universit{\"a}t M{\"u}nchen, Lichtenbergstrasse 2a, 
D-85748 Garching bei M\"unchen, Germany\\
$^{6}$ICRANet, Piazzale della Repubblica 10, I-65122, Pescara, 
Italy\\
$^{7}$Technische Universit{\"a}t M{\"u}nchen, Physik-Department, 
James-Frank-Str. 1, D-85748 Garching bei M{\"u}nchen, Germany\\
}

\date{Accepted 2019 January 19. Received 2019 January 15; in original form 2018 September 21}

\pubyear{2019}

\begin{document}
\label{firstpage}
\pagerange{\pageref{firstpage}--\pageref{lastpage}}
\maketitle

\begin{abstract}
We present evidence that TXS\,0506+056, the first plausible non-stellar
neutrino source, despite appearances, is {\it not} a blazar of the BL Lac
type but is instead a masquerading BL Lac, i.e., intrinsically a
flat-spectrum radio quasar with hidden broad lines and a standard accretion
disk. This re-classification is based on: (1) its radio and \ion{O}{II}
luminosities; (2) its emission line ratios; (3) its Eddington ratio. We
also point out that the synchrotron peak frequency of TXS\,0506+056 is 
more than two orders of magnitude larger than expected by the so-called ``blazar
sequence'', a scenario which has been assumed by some theoretical models
predicting neutrino (and cosmic-ray) emission from blazars. Finally, we
comment on the theoretical implications this re-classification has
on the location of the $\gamma$-ray emitting region and our understanding
of neutrino emission in blazars.
\end{abstract}


\section{Introduction}\label{sec:Introduction}

The IceCube Collaboration together with the multi-messenger community has
recently reported on the association of high-energy neutrinos with the
blazar TXS\,0506+056 ($z=0.3365$). This has been triggered by the detection
of a muon neutrino with most probable energy $\sim 290$\,TeV from the
direction of the blazar at the $3 - 3.5\sigma$ level \citep{icfermi}. {\it
  Fermi}-LAT observations have revealed that TXS\,0506+056 was in a flaring
state at the time of the IceCube alert, with a $0.1-300$\,GeV flux higher
by a factor $\sim 6$ than the average reported in the 3LAC catalogue
\citep{Fermi3LAC,icfermi}. Follow-up observations have led to the detection
of the source with the MAGIC and VERITAS telescopes at energies $>$
100\,GeV, as well as in the X-ray, optical, and radio bands, by {\it
  Swift}/XRT, {\it NuSTAR}, ASAS-SN, VLA, and various other facilities
\citep{icfermi,veritas18,Alb18,Ansoldi18}. Subsequent analysis of archival
IceCube data has revealed $13\pm5$ muon neutrinos in excess of background
expectations arriving from the same region of the sky on a time-scale of
$4-5$ months in 2014--2015, which constitutes $3.5\sigma$ evidence for
neutrino emission from the direction of TXS\,0506+056 \citep{iconly}. A
dissection of the region around the neutrino position has further shown
that TXS 0506+056 is the only counterpart of all the neutrino emission in
the region and therefore the most plausible first high-energy neutrino
source \citep{Padovani_2018}.

From the optical spectroscopy viewpoint blazars are historically divided in two classes,
namely flat-spectrum radio quasars (FSRQs) and BL Lac objects (henceforth,
BL Lacs), with the former displaying strong, broad emission lines just like
standard quasars, and the latter instead showing at most weak emission
lines, sometimes exhibiting absorption features, and in many cases being
completely featureless \citep{UP95}. Notably, all FSRQs are of the
low-energy peaked (LBL\footnote{Blazars are divided based on the
  rest-frame frequency of the low-energy (synchrotron) hump (\nup) into LBL
  sources (\nup~$<10^{14}$~Hz [$<$ 0.41 eV]), intermediate-
  ($10^{14}$~Hz$<$ \nup~$< 10^{15}$~Hz [0.41 eV -- 4.1 eV)], and
  high-energy (\nup~$> 10^{15}$~Hz [$>$ 4.1 eV]) peaked (IBL and HBL)
  sources respectively \citep{padgio95,Abdo_2010}.}), and in a few cases,
IBL type.

In this Letter we show that TXS\,0506+056 is not what it looks like, i.e.,
a blazar of the BL Lac type, but instead is intrinsically an FSRQ.  We also
briefly comment on the implications this might have on the theoretical
modeling of this source. We use a $\Lambda$CDM cosmology with $H_0 = 70$
km s$^{-1}$ Mpc$^{-1}$, $\Omega_{\rm m,0} = 0.3$, and $\Omega_{\Lambda,0} =
0.7$.
 
\section{The nature of TXS\,0506+056}\label{sec:nature}
The first clue about the nature of TXS\,0506+056 comes from its spectral
energy distribution (SED), which shows the double-humped structure typical
of blazars \citep[e.g.][]{Padovani_2017} with a \nup~ $\lesssim 10^{15}$
Hz, which puts it in the IBL/HBL transition region (see, e.g., Fig. 7 of
\citealt{Padovani_2018} and Fig. 4 of \citealt{icfermi}).  This is
confirmed also by its steep soft X-ray spectrum
\citep[e.g.][]{Padovani_1996} and the SED upturn at $\approx 10^{18}$ Hz.

The second clue comes from its optical spectrum \citep{Paiano_2018}, which
is characterized by a power-law continuum and, apart from faint
interstellar features, displays three extremely weak emission lines
identified as \ion{O}{II} 3727 \AA, \ion{O}{III} 5007 \AA, and \ion{N}{II}
6583 \AA, with equivalent widths (EWs\footnote{The EW of a spectral line is
  a measure of its strength and it is (roughly) defined as its flux (in
  units of erg s$^{-1}$ cm$^{-2}$) normalized by the continuum level
  underneath the line (in units of erg s$^{-1}$ cm$^{-2}$ \AA$^{-1}$). It
  is therefore measured in \AA.})  ranging between 0.05 and 0.17
\AA. TXS\,0506+056 therefore fully qualifies as a BL Lac according to the
standard empirical definition \citep[EW $< 5$
  \AA;][]{Stickel_1991,Stocke_1991}.

Why do BL Lacs have low EWs? \cite{bla78} had originally suggested that the
absence of emission lines in BL Lacs was due to a very bright,
Doppler-boosted jet continuum, which was washing out the lines (said
differently, the EW was low because the continuum was high; see
also \citealt{Georganopoulos_1998}). In the years
following that paper observations of various BL Lacs, mostly selected in
the X-ray band, showed that in many cases their optical spectrum was not
swamped by a non-thermal component, as host galaxy features were very
visible \citep{Stocke_1991}. It was then thought that most BL Lacs had {\it
  intrinsically} weak lines (i.e., the EW was low because the line was
weak). To complicate matters some objects appeared to change class 
\citep[e.g.][]{Vermeulen_1995,Pian_1999}. \cite{giommibsv1,giommibsv2} have shown that these two
possibilities are not exclusive and indeed are both viable. Therefore,
objects so far classified as BL Lacs on the basis of their {\it observed}
weak, or undetectable, emission lines belong to two {\it physically
  different} classes: intrinsically weak-lined objects, and heavily diluted
broad-lined sources, which are in reality quasars. These latter objects
have been labelled ``masquerading BL Lacs'' by \cite{giommibsv2}.  We
stress that these sources typically have relatively high powers and
\nup~values (see Fig. 10 of \citealt{giommibsv1}), which translate into
more non-thermal jet-related optical light than present in low
\nup~sources.  This implies that the emission lines are more easily
diluted, explaining their BL Lac classification. In short, ``masquerading
BL Lacs'' are the missing FSRQs with relatively high \nup.

We stress that ``real'' BL Lacs and FSRQs belong to very different physical
classes, namely objects {\it without} and {\it with} high-excitation
emission lines in their optical spectra, referred to as low-excitation
(LEGs) and high-excitation galaxies (HEGs), respectively. As discussed by
\cite{Padovani_2017} the LEG/HEG classification applies to AGN in general:
quasars and Seyferts belong to the HEG category, while low-ionization
nuclear emission-line regions (LINERs) and absorption line galaxies are
classified as LEGs.

\begin{table}
 \caption{Masquerading BL Lacs 
 ($\equiv$ high \nup~FSRQs): their properties in perspective.}
 \begin{tabular}{lccc}
       & BL Lacs  & FSRQs & Masquerading BL Lacs \\
  \hline
accretion  &  inefficient&     efficient  &   efficient\\
  &  &   & (but apparently not)\\
EW                    & $< 5~$\AA    & $> 5~$\AA & $< 5~$\AA  \\
$L/L_{\rm Edd}$ & $\lesssim 0.01$  &  $\gtrsim 0.01 $ & $\gtrsim 0.01 $\\
%  \hline
  \nup & any & $\lesssim 10^{14}$ Hz & $\gtrsim10^{14}$ Hz \\
 \hline
  \end{tabular}
 \label{tab:summary}
\end{table}

There are fundamental physical differences between these two types of AGN.
Namely, LEGs exhibit radiatively inefficient accretion related to low
Eddington ratio\footnote{This is the ratio between the (accretion-related)
  observed luminosity and the Eddington luminosity, $L_{\rm Edd} = 1.26
  \times 10^{46}~(M/10^8 \rm M_{\odot})$ erg s$^{-1}$, where $\rm
  M_{\odot}$ is one solar mass. This is the maximum isotropic luminosity a
  body can achieve when there is balance between radiation pressure (on the
  electrons) and gravitational force (on the protons).} ($L/L_{\rm Edd}
\lesssim 0.01$), while HEGs accrete in a radiatively efficient manner at
high Eddington rates ($0.01 \lesssim L/L_{\rm Edd} \lesssim 1$\footnote{As
  stressed by \cite{Padovani_2017} the dividing line in $L/L_{\rm Edd}$
  needs to be considered only in a statistical sense. The fundamental
  physical separation, in fact, may be also be a function of other
  parameters (such as spin and black hole mass) and in addition one should
  keep in mind that the observational data used to constrain this
  separation are subject to measurement and computational uncertainties and
  biases.}; e.g. \citealt{Padovani_2017}). From a theoretical perspective,
the observed difference in $L/L_{\rm Edd}$ is generally associated with a
switch between a standard accretion, i.e. radiatively efficient,
geometrically thin (but optically thick) disk accretion flow
\citep{Shakura_1973} and a radiatively inefficient, geometrically thick
(but optically thin) disk accretion flow \citep[e.g.][]{Narayan_1995}.
Table \ref{tab:summary} presents an overview of the properties of
masquerading BL Lacs compared to those of real BL Lacs and FSRQs.

We contend that TXS\,0506+056 is a HEG and therefore a masquerading BL Lac,
i.e., intrinsically an FSRQ with hidden broad lines and a standard Shakura
-- Sunyaev accretion disk. This claim is based on various pieces of
evidence:

\begin{enumerate}
\item Its radio power ($P_{\rm 1.4GHz} \sim 1.8 \times 10^{26}$ W
  Hz$^{-1}$) and \ion{O}{II} luminosity ($L_{\rm \ion{O}{II}} \sim 2 \times
  10^{41}$ erg s$^{-1}$: \citealt{Paiano_2018}) put this source right in
  the middle of the locus of jetted (radio-loud) quasars \citep[Fig. 4
    of][]{Kalfountzou_2012}. Moreover, HEGs become the dominant 
    population in the radio sky above $P_{\rm 1.4GHz} \sim 10^{26}$ W Hz$^{-1}$
    \citep{hec14};
%log P_r ~ 25.75 W/Hz/sr, log L_OII ~ 34.3 W
\item The optical spectrum of TXS\,0506+056 resembles that of a Seyfert 2
  galaxy from the point of view of the emission line ratios \citep{Paiano_2018},
  which implies it is a HEG;
\item Its $L/L_{\rm Edd}$ is $> 0.01$. We estimate the black hole mass by
  assuming the host galaxy to be a typical giant elliptical with absolute
  R-band magnitude $M(R) \sim -22.9$ \citep{Paiano_2018} and then by using
  the \cite{McLure_2002} relationship between black hole mass and bulge
  $M(R)$ to derive $M_{\rm BH} \approx 3 \times 10^8 M_{\odot}$. This
  translates into $L_{\rm Edd} \approx 4 \times 10^{46}$ erg s$^{-1}$. We
  note that the true value cannot be much larger than
  this. \cite{Paiano_2018} have in fact estimated a lower limit on the
  redshift $> 0.3$ based on the lack of absorption features due to the host
  galaxy. Given the closeness of this value to the measured redshift this
  means that the host galaxy can only be at the same level, or fainter,
  than assumed, which in turns means that the black hole mass cannot be
  much higher than estimated and that the resulting $L/L_{\rm Edd}$ is a
  lower limit. As for the (thermal) bolometric luminosity, we use the
  relationships between $L_{\rm bol}$ and $L_{\rm \ion{O}{II}}$ and $L_{\rm
    \ion{O}{III}}$ \citep{Punsly_2011} to derive $L_{\rm bol} \sim 9 
    \times 10^{45}$ erg
  s$^{-1}$ and $L_{\rm bol} \sim 3 \times 10^{45}$ erg s$^{-1}$
  respectively. Since \cite{Punsly_2011} have shown that these values are
  overestimated for jetted (radio-loud) quasars because of a sizeable
  jet-induced contribution we conservatively divide the logarithmic average 
  by 3 (based on their Fig. 3), which gives $L_{\rm bol} \sim 1.7 \times 10^{45}$
  erg s$^{-1}$. We finally obtain $L/L_{\rm Edd} \gtrsim 0.04$.
\end{enumerate}

We can also estimate the broad-line region (BLR) luminosity in two
different ways (top-down and bottom-up): 1. assuming $\langle L_{\rm
  bol}/L_{\rm disk} \rangle \approx 2$, which is consistent with typical
quasar SEDs \citep[e.g.][]{Richards_2006}, we obtain $L_{\rm disk} \sim 8
\times 10^{44}$ erg s$^{-1}$, which translates into $L_{\rm BLR} \sim 8
\times 10^{43}$ erg s$^{-1}$ for a standard covering factor $\sim 10$ per
cent; 2. we derive the narrow line luminosity (NLR) from $L_{\rm NLR} = 3
\times (3 \times L_{\rm \ion{O}{II}} + 1.5 \times L_{\rm \ion{O}{III}})
\sim 3 \times 10^{42}$ erg s$^{-1}$ \citep{Rawlings_1991}, from which we
get $L_{\rm BLR} \sim 3 \times 10^{43}$ erg s$^{-1}$ assuming $L_{\rm
  BLR}/L_{\rm NLR} \sim 10$, which is typical of FSRQs
\citep{Gu_2009}. These give a logarithmic average $\sim 5 \times 10^{43}$
erg s$^{-1}$.

\cite{Ghisellini_2011} have proposed a classification scheme to divide BL
Lacs from FSRQs, which is based on $L_{\rm BLR}$ in Eddington units, and
set at a dividing value of $L_{\rm BLR}/L_{\rm Edd} \sim 5 \times 10^{-4}$,
i.e. $L_{\rm disk}/L_{\rm Edd} \sim 0.005$ (for a $\sim 10$ per cent
covering factor). This turns out to be also the value, which separates
radiatively efficient from radiatively inefficient regimes and in fact
coincides with our LEG/HEG division (since $\langle L_{\rm bol}/L_{\rm
  disk} \rangle \approx 2$). In the case of TXS\,0506+056 $L_{\rm
  BLR}/L_{\rm Edd} \sim 0.001$, which implies that this source is an FSRQ
also according to the \cite{Ghisellini_2011}
criterion. \cite{Sbarrato_2012} have proposed a further division between BL
Lacs and FSRQs at $L_{\gamma}/L_{\rm Edd} \sim 0.1$. TXS\,0506+056, with
(an average) $L_{\gamma}/L_{\rm Edd} \sim 0.7$, is well into the FRSQ
region.

\section{TXS\,0506+056 as a ``blazar sequence'' outlier}\label{sec:sequence}

\begin{figure}
%\hspace{-2.2em}
\includegraphics[width=0.5\textwidth]{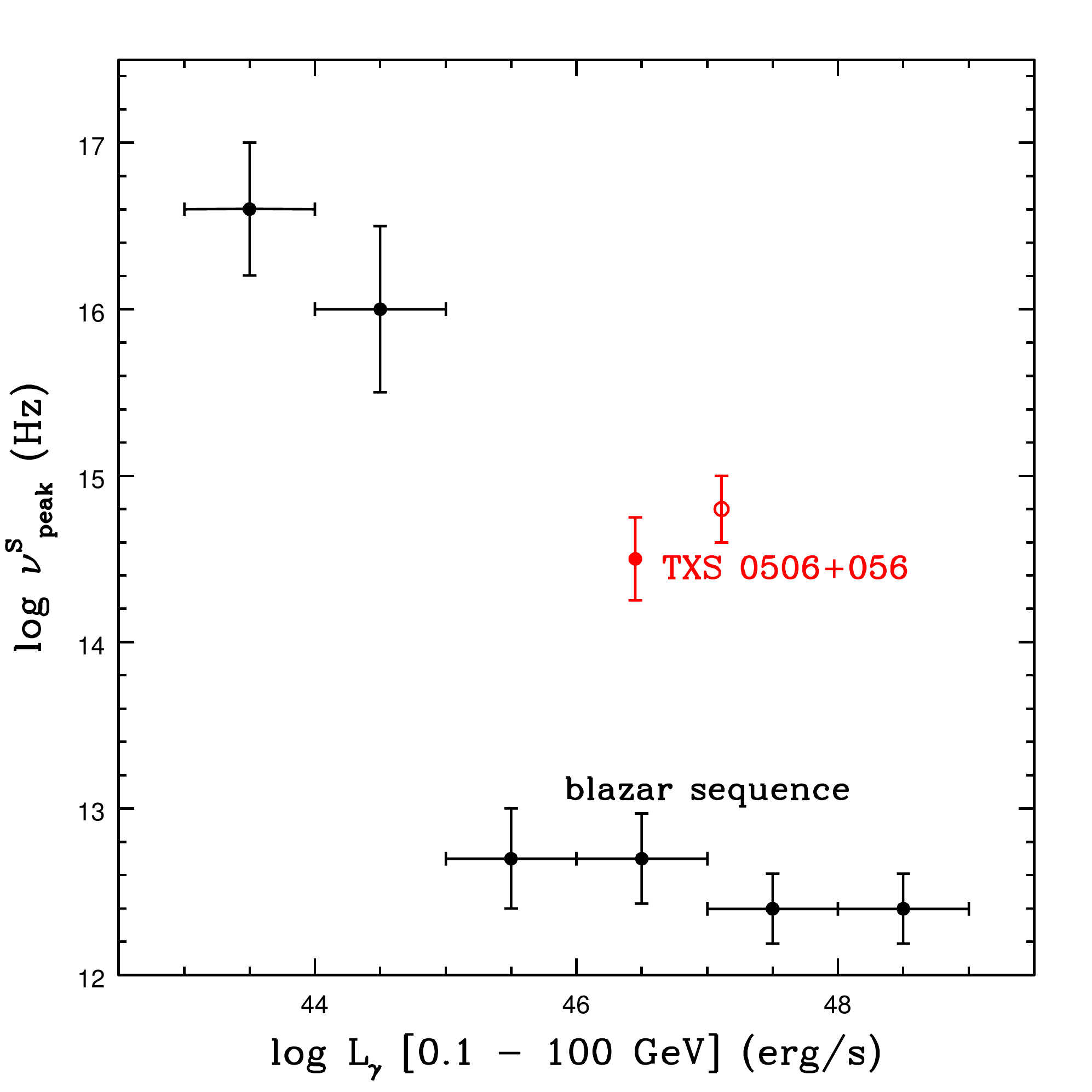}
\caption{\nup~versus $L_{\gamma}$ for the revised blazar sequence (black
  points; \citealt{Ghisellini_2017}) and TXS\,0506+056 (red points: average
  [filled] and $\gamma$-ray flare [open] values). Its $\gamma$-ray powers
  come from \protect\cite{icfermi}, while the \nup~values were derived by
  fitting the SED.  The error bars denote the sample dispersion (blazar
  sequence) and the uncertainty (TXS\,0506+056) respectively.}
\label{fig:sequence}
\end{figure}

The existence of a strong anti-correlation between bolometric luminosity
and \nup, the so-called ``blazar sequence'', has been the subject of
intense debate since it was first proposed by \cite{fossati98} and
\cite{ghis98} \citep[see discussion and references in,
  e.g.,][]{giommibsv1}. This is related to the apparent lack of FSRQs of
the HBL type (Section \ref{sec:Introduction}).  We do not wish here to
(re-)enter this controversy but simply to look at things from the point of
view of TXS\,0506+056.

Many blazar sequence outliers have been discovered so far, both in the low
power -- low \nup~\citep[e.g.][]{Padovani_2003,Anton_2005, Raiteri_2016}
and high power -- high \nup~
\citep[e.g.][]{Padovani_2012,Kaur_2017,Kaur_2018} regions of parameter
space.

We have estimated \nup~for TXS\,0506+056 by fitting both its average SED
(using all available archival data: \nup~$\sim 10^{14.5\pm0.25}$ Hz) and the
one close to the time of the IceCube-170922A neutrino alert
\citep[][\nup~$\sim 10^{14.8\pm0.2}$ Hz; see also \citealt{Keivani:2018rnh}, who 
find \nup~$\lesssim 10^{14.5}$ Hz] {Padovani_2018}. Given its
luminosities at various frequencies ($L_{\rm peak} \sim 10^{46}$ erg s$^{-1}$, 
Section \ref{sec:nature}, and
\citealt{Padovani_2018}), TXS\,0506+056 appears to be an outlier of the
blazar sequence (see, e.g., Fig. 4 of \citealt{Meyer_2011}, 
Fig. 10 of \citealt{giommibsv1}, and
Fig. 6 of \citealt{Padovani_2012}).

\cite{Ghisellini_2017} have revisited the blazar sequence by using the {\it
  Fermi} 3LAC sample \citep{Fermi3LAC}. \nup~now changes quite abruptly as
function of $L_{\gamma}$ (0.1 -- 100 GeV), as shown in
Fig. \ref{fig:sequence} (black points). The same figure shows also the
location of TXS\,0506+056 in its average state (red filled symbol) and
during the IceCube-170922A neutrino alert (red open symbol). TXS\,0506+056
is an obvious outlier even of the revised blazar sequence: given its
$L_{\gamma}$, its \nup~should be more than two orders of magnitude smaller to
fit the sequence. We note that this is not unexpected, as masquerading BL
Lacs have high powers and high \nup~and therefore are by definition
outliers.

Given that the first plausible high-energy neutrino source does not follow
the blazar sequence, theoretical models predicting neutrino (and cosmic-ray)
emission from blazars, which have the blazar sequence embedded in their
calculations \citep[e.g.][]{2014PhRvD..90b3007M,Rodrigues_2018} will need to be revised.
  
% \section{Discussion, theoretical implications, and summary}
\section{Theoretical implications and summary}
Using estimates of the $\gamma \gamma$ opacity on BLR photons and of the
photopion $p\pi$ efficiency, we now discuss the implications of the
presence of a BLR region on the location of the $\gamma$-ray emitting
region and the neutrino output, assuming that $\gamma$-rays and neutrinos
are produced in the same region of the jet.

The BLR photon field attenuates $\gamma$-rays of observed energy
$E_{\gamma}\approx \left[25\,{\rm GeV}/(1+z)\right] \left(\epsilon_{\rm
  BLR}/\epsilon_{\gamma}\right)$ \citep[e.g.,][]{GT09,2009ApJ...704...38S},
where $\epsilon_{\rm BLR} = 10.2$\,eV is the energy of the Ly$\alpha$ line,
which makes the strongest contribution to the total BLR emission
(quantities denoted with capital and lowercase letters refer to the
observer and black-hole rest frame, respectively). The FSRQ nature of
TXS\,0506+056 also implies the existence of a dusty torus. However,
attenuation of its infrared emission would become important only at
$E_{\gamma}\gtrsim 1$~TeV.

We thus calculate the optical depth for $\gamma$-rays on the BLR photons of
TXS\,0506+056 as a function of distance of the emitting region, $R_{\rm
  em}$, from the central engine using the method of \citet{BE16}. We assume
that the BLR is a spherical emitting shell with constant emissivity and
width $0.2\ R_{\rm BLR}$, extending from $0.8\, R_{\rm BLR}$ to
$R_{\rm BLR}$ (see Fig.~1 of \citealp{BE16}).  For the luminosity of the
BLR we use the value derived in Section \ref{sec:nature}. We estimate the
radius of the BLR as $R_{\rm BLR} \simeq 10^{17}\, L_{\rm d,45}^{1/2}\ {\rm
  cm} \approx 7 \times 10^{16}$\,cm \citep[][]{GT08}, where $L_{\rm d,45} =
L_{\rm disk}/(10^{45}\,{\rm erg}\,{\rm s^{-1}})$.

We model the radiation field of the BLR considering the 21 strongest lines
from \citet{Francis91}. We find that for the luminosity derived in Section
\ref{sec:nature} ($L_{\rm BLR} \approx 5\times10^{43}~{\rm erg}~{\rm
  s}^{-1}$), $\tau_{\rm BLR}(E_{\gamma} = 100\,\rm GeV) \approx 1$ at
$R_{\rm em} \approx 0.9\, R_{\rm BLR} \approx 6.5\times 10^{16}$\,cm. The
SED of TXS\,0506+056 as measured by MAGIC and VERITAS is consistent with
significant absorption above 100\,GeV and $\tau_{\gamma \gamma} (E_{\gamma}
\approx 100\,\rm GeV) \approx 1$, when compared to the non strictly
simultaneous \textit{Fermi}-LAT data \citep{Ansoldi18,veritas18}.  This
absorption is not related to the extragalactic background light, as the
optical depth for 100\,GeV photons from a source at $z \approx 0.33$ is
$\lesssim 0.1$.  If the reported opacity is attributed to the BLR alone,
then the $\gamma$-ray emitting region of TXS\,0506+056 should be located at
its outer edge, i.e., $R_{\rm em} \gtrsim 0.9 R_{\rm BLR}$. Were the
emission region closer to the central engine, stronger internal absorption
of the 100\,GeV $\gamma$-rays by the BLR would have been expected
(e.g. \citealp{Costamante:2018anp}). We note that for our assumed
dependence of $R_{\rm BLR}$ on $L_{\rm disk}$, the BLR energy density is
independent of $L_{\rm disk}$.  Thus, varying $L_{\rm disk}$ within the
uncertainty range quoted in Section \ref{sec:nature} changes the allowed
range of $R_{\rm em}$ relative to $R_{\rm BLR}$ by $\lesssim 3$ per cent.

\begin{figure}
\centering
\includegraphics[width=0.53\textwidth]{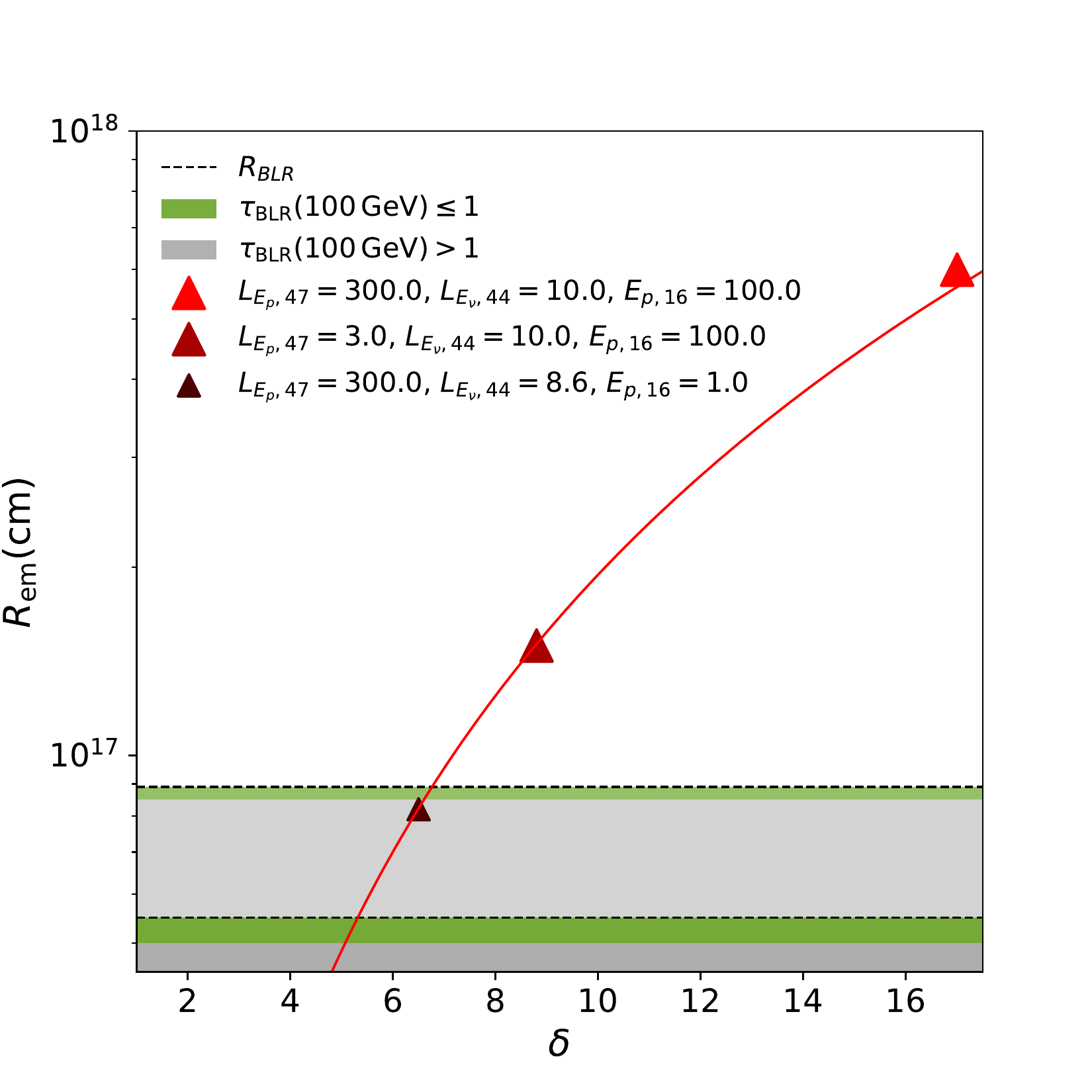}
\caption{Location of the emitting region, $R_{\rm em}$, along the jet as a
  function of the Doppler factor, $\delta$, for $t_{\rm v}=1$~day (red
  solid line) -- see eq.~(\ref{eq:Rem}). Dashed horizontal lines mark the
  values of the BLR radius that correspond to the range of $L_{\rm BLR}$
  derived in Section~2. Green bands denote the respective ranges of $R_{\rm
    em}$, where $\tau_{\rm BLR}(E_\gamma=100 \ {\rm GeV})\lesssim 1$, and
  grey bands regions where $\tau_{\rm BLR}(E_\gamma=100 \ {\rm GeV})> 1$. The
  minimum distance corresponding to $\delta_{\min}$ (see
  eq.~(\ref{eq:dmin})) is also indicated with red symbols for different
  parameters marked on the plot. The size of the symbols is proportional to
  the predicted all-flavour bolometric neutrino luminosity (see legend;
  where we used the notation $Q_{x} = Q/10^x$ in erg s$^{-1}$, and eV, for
  $L_{\rm E_{\rm p}},L_{\rm E_{\nu}}$, and $E_{\rm p}$ respectively). If the blob lies within
  the BLR, then the constraint on $\delta_{\min}$ (eq.~(\ref{eq:dmin}))
  must be modified to include interactions on BLR photons. Still,
  $E_{\nu}L_{\rm E_{\nu}}\lesssim 10^{45}$~erg s$^{-1}$ -- see
  eq.~(\ref{eq:Lv}). }
\label{fig:doppler}
\end{figure}

We can relate the radius of the emission region (blob), $R^{\prime}_{\rm
  b}$, to the distance of the dissipation region from the central engine
$R_{\rm em}$, by assuming that the blob covers the cross-sectional area of
the jet:
\begin{equation}
R_{\rm em} = \tan \theta_j ^{-1} R_{\rm b}^{\prime} \approx \theta_j^{-1}R_{\rm b}^{\prime} \approx \Gamma R_{\rm b}^{\prime}\approx \delta^2 c t_{\rm v}(1+z)^{-1},
\label{eq:Rem}
\end{equation}
where $\theta_j\approx1/\Gamma$ is the half-opening angle of a conical jet,
$\Gamma$ is the Lorentz factor of the bulk flow, $\delta \approx \Gamma$ is
the Doppler factor, $t_{\rm v}$ is the observed variability timescale, and
primed variables denote quantities in the co-moving frame of the blob. For
a fixed $t_{\rm v}$ (assumed equal to 1 day based on X-ray variability:
\citealt{Keivani:2018rnh}), different values of $\delta$ correspond to
different locations according to eq.~(\ref{eq:Rem}), as shown in
Fig.~\ref{fig:doppler} (red solid line). The range of $R_{\rm em}$ where
$\tau_{\rm BLR}(E_{\gamma} = 100\,\rm GeV) \lesssim 1$ (green bands) lies
within the outer radius of the BLR (dashed lines).

We next discuss constraints due to the electromagnetic (EM) cascade
emission produced in the source from photohadronic ($p\gamma$)
interactions. We first derive constraints due to interactions with the
blazar synchrotron photons (see also numerical results by 
\citealp{2018NatAs.tmp..154G,2019MNRAS.483L..12C}). We then 
discuss how the presence of the BLR changes these results.
 
We require that the luminosity of the photon component produced by
$p\gamma$ interactions does not exceed the luminosity in the X-ray band
($0.3 - 10$ keV), i.e. $L_{\rm X,lim}\simeq 3\times 10^{44}\, F_{\rm X,lim,
  -12}$~erg~s$^{-1}$ \citep{icfermi, Keivani:2018rnh}. Using eqs.~(9),
(13), and (14) of \cite{MOP18} and assuming that a fraction $f_x$ of the
bolometric cascade luminosity will emerge in the X-ray band, we derive a
lower limit on the Doppler factor of the emitting region:
\begin{eqnarray}
\delta_{\min}^{2+2\beta} \approx f(\beta)
\frac{3\hat{\sigma}_{\rm 
p\pi}L_{\rm s}}{4 \pi c^2 t_{\rm v}E_{\rm s}}\left[\frac{2E_{\rm p}E_{\rm s}(1+z)^2}{m_{\rm p}c^2 \bar{\epsilon}_{\Delta}}\right]^{\beta-1}\frac{E_{\rm 
p}L_{\rm E_{\rm p}}}{f_x^{-1} L_{\rm X, lim}},
\label{eq:dmin}
\end{eqnarray}
where $\beta\sim 2.5$ is the photon index in the {\it Swift}/XRT energy
band \citep{icfermi, Keivani:2018rnh}, $f(\beta)=[2/(1+\beta)][5/16 +
  g(\beta)/2]$, $g(\beta)\simeq 0.01\,(30)^{\beta-1}$, $\hat{\sigma}_{\rm
  p\pi}\simeq 7\times10^{-29}$~cm$^2$, $\bar{\epsilon}_{\Delta}\sim
0.3$~GeV, $L_{\rm s}\sim 10^{46}$~erg s$^{-1}$ and $E_{\rm s}\sim 1$~eV are
the observed (isotropic) synchrotron luminosity and photon energy,
respectively. The only free parameters in eq.~(\ref{eq:dmin}) are the
proton energy $E_{\rm p}$ and isotropic proton luminosity $E_{\rm
  p}L_{\rm E_{\rm p}}$. For $\delta=\delta_{\min}$ and $f_x \sim 0.1$ we obtain
an upper limit on the all-flavour neutrino luminosity
\citep[e.g.][]{MOP18}:
\begin{eqnarray}
\label{eq:Lv}
E_\nu L_{\rm E_{\nu}}\lesssim \frac{3}{8}\left[\frac{5}{16}+\frac{g(\beta)}{2}\right]^{-1}\frac{L_{\rm X,lim}}{f_x} \approx 10^{45} \frac{L_{\rm X,lim, 44.5}}{f_{x,-1}} \ {\rm erg \ s}^{-1},
\end{eqnarray}
in agreement with detailed numerical modeling of the flaring SED of
TXS~0506+056 \citep[e.g.][]{Keivani:2018rnh}.

The location of a blob moving with $\delta_{\min}$ can be estimated by
substitution of eq.~ (\ref{eq:dmin}) into eq.~(\ref{eq:Rem}). The results
obtained for different values of $E_{\rm p}$ and $E_{\rm p}L_{\rm E_{\rm p}}$
are presented in Fig.~\ref{fig:doppler} (red symbols). We find that the
blob is located beyond the BLR (dashed horizontal lines) unless $E_{\rm
  p}\le10$~PeV or $E_{\rm p}L_{\rm E_{\rm p}}<10^{47}$~erg s$^{-1}$. In the
latter case, the BLR provides additional target photons for $p\gamma$
interactions and one might expect that a higher neutrino luminosity can be
achieved. However, the accompanying EM cascade component would be
accordingly more luminous \citep[see also][]{PM15}. By requiring that the
EM cascade emission from $p\gamma$ interactions on BLR photons does not
overshoot the X-ray data (see also \citealt{petro_17}), we can also derive
an upper bound on the expected $E_\nu L_{\rm E_\nu}$ from a blob located within
the BLR, which is still given by eq.~(\ref{eq:Lv}).

In conclusion, $\gamma \gamma$ opacity constraints allow the emitting
region to be at the outer edge of the BLR (see Fig.~\ref{fig:doppler}), but
the constraints from the cascade emission place an upper bound of $\sim
10^{45}$~erg s$^{-1}$ to the all-flavour neutrino luminosity produced
within a blob, irrespective of its location with respect to the BLR. We
notice that the upper bound derived here is consistent with\footnote{In an
  ensemble of faint sources with a summed expectation of order 1, one might
  observe a neutrino even if the individual expectation value is $\ll 1$
  \citep{icfermi}.  Hence the upper limits.}, and well below, the neutrino
luminosity implied by the IceCube observations, assuming that the source
was emitting neutrinos throughout the whole IceCube observation period of
7.5 years ($\lesssim 10^{46}$ erg s$^{-1}$) and 6 months ($\lesssim 2
\times 10^{47}$ erg s$^{-1}$) respectively \citep{icfermi}. The explanation
of neutrino luminosities as high as $\sim 10^{47}$~erg s$^{-1}$ as seen
during the 2014--2015 neutrino flare \citep{Padovani_2018,iconly} requires
more complex theoretical scenarios that invoke more than one emitting
region \citep[e.g.][]{MOP18}.

In summary, the radio and \ion{O}{II} luminosities, emission line ratios,
and Eddington ratio of TXS\,0506+056, all point to its re-classification as
a masquerading BL Lac, namely an FSRQ with the emission lines heavily
diluted by a strong, Doppler-boosted jet. Moreover, TXS\,0506+056 
has a \nup, which is more than two orders of magnitude larger than expected by
the blazar sequence. These two facts are likely to have an
impact on the theoretical modeling of this source and on our understanding
of neutrino emission in blazars.

\section*{Acknowledgments}
PP thanks the ASI Science Data Center (SSDC) for the
hospitality and partial financial support for his visit. PG acknowledges
the support of the TUM - IAS, funded by the German Excellence Initiative (and the
European Union Seventh Framework Programme under grant agreement
no. 291763). MP acknowledges support by the Lyman Jr. Spitzer Postdoctoral
Fellowship. FO acknowledges useful corrrspondence and discussions with
Markus B\"{o}ttcher, Paul Els, Konstancja Satalecka, and Michael
Unger. This work is supported by the Deutsche Forschungsgemeinschaft
through grant SFB\,1258 ``Neutrinos and Dark Matter in Astro- and Particle
Physics''.

\label{lastpage}

% Don't change these lines
\bsp	% typesetting comment


\begin{thebibliography}{}
\bibitem[\protect\citeauthoryear{Abdo et al.}{2010}]{Abdo_2010} Abdo A. A.,
  Ackermann M., Ajello M., et al., 2010, ApJ, 716, 30
\bibitem[\protect\citeauthoryear{Abeysekara et al.}{2018}]{veritas18}
  Abeysekara A.~U., et al., 2018, ApJ, 861, L20
\bibitem[\protect\citeauthoryear{Ackermann et al.}{2015}]{Fermi3LAC}
  Ackermann M., et al., 2015, ApJ, 810, 14
\bibitem[\protect\citeauthoryear{Albert et al.}{2018}]{Alb18} Albert A., et
  al., 2018, ApJ, 863, L30
\bibitem[\protect\citeauthoryear{Ansoldi et al.}{2018}]{Ansoldi18} Ansoldi
  S., et al., 2018, ApJ, 863, L10
\bibitem[\protect\citeauthoryear{Ant{\'o}n \& Browne}{2005}]{Anton_2005}
  Ant{\'o}n S., Browne I.~W.~A., 2005, MNRAS, 356, 225
\bibitem[\protect\citeauthoryear{Blandford \& Rees}{1978}]{bla78} Blandford
  R.~D., Rees M. J., 1978, in Pittsburg Conference on BL Lac Objects,
  Ed. A.~M. Wolfe, Pittsburgh, University of Pittsburgh press, p. 328
\bibitem[\protect\citeauthoryear{B{\"o}ttcher \& Els}{2016}]{BE16}
  B{\"o}ttcher M., Els P., 2016, ApJ, 821, 102
  \bibitem[Cerruti et al.(2019)]{2019MNRAS.483L..12C} 
  Cerruti, M., Zech, A., Boisson, C., et al.\ 2019, \mnras, 483, L12  
\bibitem[\protect\citeauthoryear{Costamante et
    al.}{2018}]{Costamante:2018anp} Costamante L., Cutini S., Tosti G.,
  Antolini E., Tramacere A., 2018, MNRAS, 477, 4749
\bibitem[\protect\citeauthoryear{Fossati et al.}{1998}]{fossati98} Fossati
  G., Maraschi L., Celotti A., Comastri A., Ghisellini G., 1998 MNRAS, 299,
  433
\bibitem[\protect\citeauthoryear{Francis et al.}{1991}]{Francis91} Francis
  P.~J., Hewett P.~C., Foltz C.~B., Chaffee F.~H., Weymann R.~J., Morris
  S.~L., 1991, ApJ, 373, 465
\bibitem[\protect\citeauthoryear{Gao, Fedynitch, Winter \& Pohl}{2018}]
  {2018NatAs.tmp..154G} Gao S., Fedynitch A., Winter W., Pohl M., 2018, Nature Astronomy, 154 
\bibitem[\protect\citeauthoryear{Georganopoulos \& Marscher}{1998}]{Georganopoulos_1998} 
Georganopoulos M., Marscher A.~P., 1998, ApJ, 506, 621 
\bibitem[\protect\citeauthoryear{Ghisellini et al.}{1998}]{ghis98}
  Ghisellini G., Celotti A., Fossati G., Maraschi L., Comastri A., 1998,
  MNRAS, 301, 451
\bibitem[\protect\citeauthoryear{Ghisellini \& Tavecchio}{2008}]{GT08}
  Ghisellini G., Tavecchio F., 2008, MNRAS, 387, 1669
\bibitem[\protect\citeauthoryear{Ghisellini \& Tavecchio}{2009}]{GT09}
  Ghisellini G., Tavecchio F., 2009, MNRAS, 397, 985
\bibitem[\protect\citeauthoryear{Ghisellini et al.}{2011}]{Ghisellini_2011}
  Ghisellini G., Tavecchio F., Foschini L., Ghirlanda G., 2011, MNRAS, 414,
  2674
\bibitem[\protect\citeauthoryear{Ghisellini et al.}{2017}]{Ghisellini_2017}
  Ghisellini G., Righi C., Costamante L., Tavecchio F., 2017, MNRAS, 469,
  255
\bibitem[\protect\citeauthoryear{Giommi et al.}{2012}]{giommibsv1} Giommi
  P., Padovani P., Polenta G., Turriziani S., D'Elia V., Piranomonte S.,
  2012, MNRAS, 420, 2899
\bibitem[\protect\citeauthoryear{Giommi et al.}{2013}]{giommibsv2} Giommi
  P., Padovani P. \& Polenta G., 2013, MNRAS, 431, 1914  
\bibitem[\protect\citeauthoryear{Gu, Chen, \& Cao}{2009}]{Gu_2009} Gu M.,
  Chen Z., Cao X., 2009, MNRAS, 397, 1705
\bibitem[\protect\citeauthoryear{Heckman \& Best}{2014}]{hec14} Heckman
  T.~M., Best P.~N., 2014, ARA\&A, 52, 589  
\bibitem[\protect\citeauthoryear{IceCube Collaboration}{2018}]{iconly}
  IceCube Collaboration, 2018, Science, 361, 147
\bibitem[\protect\citeauthoryear{IceCube Collaboration et al.}{2018}]{icfermi}
  IceCube Collaboration, Fermi-LAT, MAGIC, AGILE, ASAS-SN, HAWC, H.E.S.S.,
  INTEGRAL, Kanata, Kiso, Kapteyn, Liverpool Telescope, Subaru,
  Swift/NuSTAR, VERITAS, VLA/17B-403 teams, 2018, Science, 361, eaat1378
\bibitem[\protect\citeauthoryear{Kalfountzou et
    al.}{2012}]{Kalfountzou_2012} Kalfountzou E., Jarvis M.~J., Bonfield
  D.~G., Hardcastle M.~J., 2012, MNRAS, 427, 2401
\bibitem[\protect\citeauthoryear{Kaur et al.}{2017}]{Kaur_2017} Kaur A., et
  al., 2017, ApJ, 834, 41
\bibitem[\protect\citeauthoryear{Kaur et al.}{2018}]{Kaur_2018} Kaur A.,
  Rau A., Ajello M., Dom{\'{\i}}nguez A., Paliya V.~S., Greiner J.,
  Hartmann D.~H., Schady P., 2018, ApJ, 859, 80
\bibitem[\protect\citeauthoryear{Keivani et al.}{2018}]{Keivani:2018rnh}
  Keivani A., et al., 2018, ApJ, 864, 84
\bibitem[\protect\citeauthoryear{McLure \& Dunlop}{2002}]{McLure_2002}
  McLure R.~J., Dunlop J.~S., 2002, MNRAS, 331, 795
\bibitem[\protect\citeauthoryear{Meyer et al.}{2011}]{Meyer_2011} 
Meyer E.~T., Fossati G., Georganopoulos M., Lister M.~L., 2011, ApJ, 740, 98 
\bibitem[\protect\citeauthoryear{Murase, Inoue, \& Dermer}{2014}]{2014PhRvD..90b3007M} 
Murase K., Inoue Y., Dermer C.~D., 2014, PhRvD, 90, 023007 
\bibitem[\protect\citeauthoryear{Murase, Oikonomou, \&
    Petropoulou}{2018}]{MOP18} Murase K., Oikonomou F., Petropoulou M.,
  2018, ApJ, 865, 124 
\bibitem[\protect\citeauthoryear{Narayan \& Yi}{1995}]{Narayan_1995}
  Narayan R., Yi I., 1995, ApJ, 452, 710
\bibitem[\protect\citeauthoryear{Padovani \& Giommi}{1995}]{padgio95}
  Padovani P., Giommi P., 1995, ApJ, 444, 567  
\bibitem[\protect\citeauthoryear{Padovani \& Giommi}{1996}]{Padovani_1996}
  Padovani P., Giommi P., 1996, MNRAS, 279, 526
\bibitem[\protect\citeauthoryear{Padovani et al.}{2003}]{Padovani_2003}
  Padovani P., Perlman E.~S., Landt H., Giommi P., Perri M., 2003, ApJ,
  588, 128
\bibitem[\protect\citeauthoryear{Padovani, Giommi, \&
    Rau}{2012}]{Padovani_2012} Padovani P., Giommi P., Rau A., 2012, MNRAS,
  422, L48
\bibitem[\protect\citeauthoryear{Padovani et al.}{2017}]{Padovani_2017}
  Padovani P., et al., 2017, A\&ARv, 25, 2
\bibitem[\protect\citeauthoryear{Padovani et al.}{2018}]{Padovani_2018}
  Padovani P., Giommi P., Resconi E., T. Glauch, B. Arsioli, N. Sahakyan,
  M. Huber, 2018, MNRAS, 480, 192
\bibitem[\protect\citeauthoryear{Paiano et al.}{2018}]{Paiano_2018} Paiano
  S., Falomo R., Treves A., Scarpa R., 2018, ApJ, 854, L32
\bibitem[\protect\citeauthoryear{Petropoulou \& Mastichiadis}{2015}]{PM15}
  Petropoulou M., \& Mastichiadis A., 2015, MNRAS, 447, 36
\bibitem[\protect\citeauthoryear{Petropoulou et al.}{2017}]{petro_17}
  Petropoulou M., Nalewajko K., Hayashida M., Mastichiadis A., 2017, MNRAS,
  467, L16
\bibitem[\protect\citeauthoryear{Pian et al.}{1999}]{Pian_1999} 
Pian E., et al., 1999, ApJ, 521, 112   
\bibitem[\protect\citeauthoryear{Punsly \& Zhang}{2011}]{Punsly_2011}
  Punsly B., Zhang S., 2011, MNRAS, 412, L123
\bibitem[\protect\citeauthoryear{Raiteri \& Capetti}{2016}]{Raiteri_2016}
  Raiteri C.~M., Capetti A., 2016, A\&A, 587, A8
\bibitem[\protect\citeauthoryear{Rawlings \&
    Saunders}{1991}]{Rawlings_1991} Rawlings S., Saunders R., 1991, Nature,
  349, 138
\bibitem[\protect\citeauthoryear{Richards et al.}{2006}]{Richards_2006}
  Richards G.~T., et al., 2006, ApJS, 166, 470
\bibitem[\protect\citeauthoryear{Rodrigues et al.}{2018}]{Rodrigues_2018}
  Rodrigues X., Fedynitch A., Gao S., Boncioli D., Winter W., 2018, ApJ,
  854, 54
\bibitem[\protect\citeauthoryear{Sbarrato et al.}{2012}]{Sbarrato_2012}
  Sbarrato T., Ghisellini G., Maraschi L., Colpi M., 2012, MNRAS, 421, 1764
\bibitem[\protect\citeauthoryear{Shakura \& Sunyaev}{1973}]{Shakura_1973}
  Shakura N.~I., Sunyaev R.~A., 1973, A\&A, 24, 337
\bibitem[\protect\citeauthoryear{Sikora et al.}{2009}]{2009ApJ...704...38S}
  Sikora M., Stawarz {\L}., Moderski R., Nalewajko K., Madejski G.~M.,
  2009, ApJ, 704, 38
\bibitem[\protect\citeauthoryear{Stickel et al.}{1991}]{Stickel_1991}
  Stickel M., Padovani P., Urry C. M., Fried J. W., K\"uhr H., 1991, ApJ,
  374, 431
\bibitem[\protect\citeauthoryear{Stocke et al.}{1991}]{Stocke_1991} Stocke
  J.~T., Morris S.~L., Gioia I.~M., Maccacaro T., Schild R., Wolter A.,
  Fleming T.~A., Henry J.~P., 1991, ApJS, 76, 813
\bibitem[\protect\citeauthoryear{Urry \& Padovani}{1995}]{UP95} Urry C.~M.,
  Padovani P., 1995, PASP, 107, 803
\bibitem[\protect\citeauthoryear{Vermeulen et al.}{1995}]{Vermeulen_1995} 
Vermeulen R.~C., Ogle P.~M., Tran H.~D., Browne I.~W.~A., Cohen M.~H., 
Readhead A.~C.~S., Taylor G.~B., Goodrich R.~W., 1995, ApJ, 452, L5 
  
\end{thebibliography}
\end{document}